\documentclass[twocolumn]{aastex631} 

\usepackage{amsmath}
\usepackage{pdfrender}
\usepackage[utf8]{inputenc} 
\usepackage{graphicx,color}
\usepackage{ulem,soul}

\newcommand{\mstar}{M_{\star}}
\newcommand{\rstar}{R_{\star}}
\newcommand{\mh}{M_{\rm h}}

\newcommand{\rt}{R_{\rm t}}

\def\msun{\, \mathrm{M}_{\hbox{$\odot$}}}
\def\rsun{\, \mathrm{R}_{\hbox{$\odot$}}}

\newcommand*{\boldgreek}[1]{%
  \textpdfrender{%
    TextRenderingMode=FillStroke,%
    LineWidth=.35pt,%
  }{#1}%
}
\newcommand{\vect}[1]{\mathbf{#1}}

\newcommand{\be}{\begin{equation}}
\newcommand{\ee}{\end{equation}}

\received{XXX}
\revised{YYY}
\accepted{ZZZ}

\submitjournal{ApJ Letters}

\shorttitle{Full stream evolution in TDEs}
\shortauthors{Bonnerot, Pessah \& Lu}

\begin{document}

\title{From Pericenter and Back: Full Debris Stream Evolution in Tidal Disruption Events}

\correspondingauthor{Clément Bonnerot}
\email{clement.bonnerot@nbi.ku.dk}

\author[0000-0001-9970-2843]{Clément Bonnerot}
\affiliation{Niels Bohr International Academy, Niels Bohr Institute, Blegdamsvej 17, DK-2100 Copenhagen Ø, Denmark\\}

\author[0000-0001-8716-3563]{Martin E. Pessah}
\affiliation{Niels Bohr International Academy, Niels Bohr Institute, Blegdamsvej 17, DK-2100 Copenhagen Ø, Denmark\\}

\author[0000-0002-1568-7461]{Wenbin Lu}
\affiliation{Department of Astrophysical Sciences, Princeton University, NJ 08544, USA\\}




\begin{abstract}

When a star passes too close to a supermassive black hole, it gets disrupted by strong tidal forces. The stellar debris then evolves into an elongated stream of gas that partly falls back towards the black hole. We present an analytical model describing for the first time the full stream evolution during such a tidal disruption event (TDE). Our framework consists in dividing the stream into different sections of elliptical geometry, whose properties are independently evolved in their co-moving frame under the tidal, pressure, and self-gravity forces. Through an explicit treatment of the tidal force and the inclusion of the gas angular momentum, we can accurately follow the stream evolution near pericenter. Our model evolves the longitudinal stream stretching and both transverse widths simultaneously. For the latter, we identify two regimes depending on whether the dynamics is entirely dominated by the tidal force (ballistic regime) or additionally influenced by pressure and self-gravity (hydrostatic regime). We find that the stream undergoes transverse collapses both shortly after the stellar disruption and upon its return near the black hole, at specific locations determined by the regime of evolution considered. The stream evolution predicted by our model can be used to determine the subsequent interactions experienced by this gas that are at the origin of most of the electromagnetic emission from TDEs. Our results suggest that the accretion disk may be fed at a rate that differs from the standard fallback rate, which would provide novel observational signatures dependent on black hole spin.

\end{abstract}



\vspace{0.7cm}

\section{Introduction}
\label{sec:intro}

Encounters between stars in galactic nuclei occasionally launch one of them on a plunging near-parabolic trajectory that leads to its disruption by the central supermassive black hole \citep{rees1988}. The stellar debris then evolves into an elongated stream of gas, half of which falls back close to the compact object where shocks and accretion occur. A powerful electromagnetic signal is then emitted that constitutes a unique probe of these otherwise quiescent black holes. Such tidal disruption events (TDEs) have been observed on multiple occasions \citep[e.g.][]{van_velzen2021,sazonov2021} and the number of detections is about to skyrocket with upcoming facilities such as the Rubin Observatory \citep{Bricman2020}.

In this Letter, we present an analytical model that can follow for the first time the entire stream evolution in a TDE. It consists in dividing the stream into different sections of elliptical geometry, which are evolved in their co-moving frame under the tidal, pressure, and self-gravity forces. By including the gas angular momentum and explicitly treating the influence of the tidal force, we are able to accurately capture the hydrodynamics near the black hole. In particular, our model can follow the return of the stream to pericenter, which is not computationally feasible with current global numerical simulations.

\section{Stream evolution model}
\label{sec:model}

\subsection{Framework}

\begin{figure*}
\centering
\includegraphics[width=\textwidth]{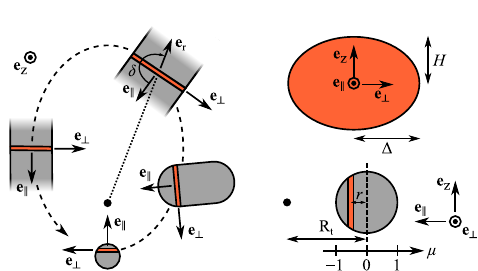}
\caption{Sketch showing the evolution of a bound section of stream (orange surface) specified by its boundness $\mu <0$ as it follows an elliptical trajectory (dashed line) around the black hole (black circle) starting from its initial location inside the star at the tidal radius (pericenter). This section remains orthogonal at all times to the unit vector $\vect{e}_{\parallel}$ locally aligned with the longitudinal direction of stream elongation. The unit vectors $\vect{e}_{\perp}$ and $\vect{e}_{\rm z}$ point along the transverse directions aligned and orthogonal to the stellar orbital plane, respectively. The transverse widths of the stream section
are given by $\Delta$ and $H$.}
\label{fig:sketch}
\end{figure*}

We consider a star approaching the black hole on a parabolic orbit with a pericenter distance equal to the tidal radius $\rt = \rstar (\mh/\mstar)^{1/3}$. Here, $\mh$ is the black hole mass while $\mstar$ and $\rstar$ denote the stellar mass and radius, respectively. We adopt the frozen-in approximation where the star is assumed to remain unaffected until it reaches the tidal radius \citep{lacy1982,rees1988}.
We divide the star into sections orthogonal to the orbital plane and located at different net distances $r = \mu \rstar$ from the stellar center, where $-1 \leq \mu \leq 1$ as schematically depicted in Fig. \ref{fig:sketch}. The centers of mass of these sections then follow a range of trajectories with the same pericenter as the star but different orbital energies $\epsilon = \mu \Delta \epsilon$ with $\Delta \epsilon = G \mh \rstar / \rt^2$ \citep{stone2013}. Accordingly, our model consists in studying the evolution of these sections, which are specified by the value of their ``boundness" $\mu$.

The centers of mass of each section evolve on different Keplerian trajectories with a position with respect to the black hole specified by $\vect{R}(\mu,t) = R \vect{e}_{\rm r}$, where $\vect{e}_{\rm r}$ is the radial unit vector. We define $\boldgreek{\ell} = -\partial \vect{R} / \partial \mu$, which is tangential to the line joining the centers of mass of all sections. This allows us to specify the orientation of each section, by requiring that it remains orthogonal to the unit vector $\vect{e}_{\parallel}\equiv \boldgreek{\ell} / \ell$, where $\ell = |\boldgreek{\ell}|$ measures the longitudinal length. The orange surface sketched in Fig. \ref{fig:sketch} depicts one of these sections with $\mu<0$ at different times as it follows a bound elliptical trajectory. The vector $\vect{e}_{\parallel}$ initially points towards the black hole and it subsequently rotates to indicate the local direction of stream elongation, as dictated by the tidal force. Based on this orientation, we define two unit vectors $\vect{e}_{\rm z}$ and $\vect{e}_{\perp} = \vect{e}_{\rm z} \times \vect{e}_{\parallel}$ that are orthogonal to and aligned with the stellar orbital plane, respectively. Along these directions, each stream section is assumed to have an elliptical geometry with vertical and in-plane transverse widths $H$ and $\Delta$ (see Fig. \ref{fig:sketch}).

\subsection{Dynamical equations}

We first describe the orientation of each section, as determined by the vector $\boldgreek{\ell} = \ell \vect{e}_{\parallel}$. Because of the tidal force, it evolves according to
\be
\boldgreek{\ddot{\ell}} = - \frac{G \mh}{R^3}\left(\boldgreek{\ell} - 3 \frac{\vect{R}\cdot \boldgreek{\ell}}{R^2} \vect{R}\right),
\label{eq:ellddot}
\ee
where dotted variables represent derivatives with respect to time. It is convenient to define the vector $\boldgreek{\Gamma} \equiv \boldgreek{\dot{\ell}} / \ell$ which, according to equation \eqref{eq:ellddot}, evolves as
\be
\dot{\boldgreek{\Gamma}} = -\frac{G\mh}{R^3}(\vect{e}_{\parallel} - 3 \cos \delta \, \vect{e}_{\rm r}) - \lambda \boldgreek{\Gamma}.
\label{eq:gammaddot}
\ee
This vector can be decomposed as $\boldgreek{\Gamma}= \lambda \vect{e}_{\parallel} + \Omega \vect{e}_{\perp}$, where $\lambda = \dot{\ell}/\ell = \boldgreek{\dot{\ell}}\cdot \vect{e}_{\parallel} / \ell$ and $\Omega = \boldgreek{\dot{\ell}}\cdot \vect{e}_{\perp} / \ell$ are the rates of elongation and rotation of the section, respectively. Projecting equation \eqref{eq:gammaddot} onto $\vect{e}_{\parallel}$ and $\vect{e}_{\perp}$ we obtain
\be
\dot{\lambda} = \Omega^2-\lambda^2-\frac{G\mh}{R^3}(1-3 \cos^2 \delta),
\label{eq:lambdadot}
\ee
\vspace{-0.35cm}
\be
\dot{\Omega} = -2 \lambda \Omega + 3\frac{G\mh}{R^3}\cos \delta \sin \delta.
\label{eq:omegadot}
\ee
Here, the angle $\delta<0$ is measured between the unit vectors $\vect{e}_{\rm r}$ and $\vect{e}_{\parallel}$ (see Fig. \ref{fig:sketch}).

The evolution of the two transverse stream widths $H$ and $\Delta$ is specified by the tidal, pressure, and self-gravity forces according to
\be
\ddot{H} = \ddot{H}_{\rm t} + \ddot{H}_{\rm p} + \ddot{H}_{\rm g},
\label{eq:hddot}
\ee
\vspace{-0.35cm}
\be
\ddot{\Delta} = \ddot{\Delta}_{\rm t} + \ddot{\Delta}_{\rm p} + \ddot{\Delta}_{\rm g}.
\label{eq:deltaddot}
\ee
The contribution from the tidal force is given by \citep{bonnerot2021-nozzle}
\be
\ddot{H}_{\rm t} = -\frac{G \mh}{R^3} H ,
\label{eq:hddott}
\ee
\vspace{-0.35cm}
\be
\ddot{\Delta}_{\rm t} = -\frac{G \mh}{R^3} \Delta (1-3 \sin^2 \delta) + \Delta \Omega^2 - 2 V \Omega,
\label{eq:detladdott}
\ee
\be
\dot{v}_{\parallel} = 3 \frac{G \mh}{R^3} \Delta \cos \delta \sin \delta + \dot{\Delta} \Omega - V \lambda,
\label{eq:vparadot}
\ee
where $V= \Delta \Omega + v_{\parallel}$. The last two terms in equation \eqref{eq:detladdott} reflect the fact that each section remains orthogonal to the direction $\vect{e}_{\parallel}$ of stream elongation. These terms compensate for shearing that imposes the gas at $\Delta \neq 0$ to move along $\vect{e}_{\parallel}$ with respect to the center of mass at the velocity $v_{\parallel}$ determined by equation \eqref{eq:vparadot}.

Following \citet{kochanek1994}, the contributions from pressure forces and self-gravity are estimated as
\be
\ddot{H}_{\rm p} = \frac{P}{\rho H},
\label{eq:hddotp}
\ee
\be
\ddot{\Delta}_{\rm p} = \frac{P}{\rho\Delta},
\label{eq:deltaddotp}
\ee
\be
\ddot{H}_{\rm g} = \ddot{\Delta}_{\rm g} = -4 \pi G \rho \frac{H \Delta}{H+\Delta}.
\label{eq:hdeltaddotg}
\ee
Equations \eqref{eq:hddotp} and \eqref{eq:deltaddotp} approximate the pressure gradients by neglecting order unity factors related to the exact transverse profiles while equation \eqref{eq:hdeltaddotg} makes use of Gauss's theorem, estimating the circumference of the elliptical section as $\pi (H+\Delta)$.

The pressure is evaluated from $P = K \rho^{5/3}$ where $K$ is a constant, as appropriate for an adiabatic evolution. The density of a given section of stream is calculated from $\rho = \Lambda/(\pi H \Delta)$, where the linear density evolves as 
\be
\dot{\Lambda} = -\lambda \Lambda,
\label{eq:rholindot}
\ee
due to longitudinal stretching at a rate obtained by solving equation \eqref{eq:lambdadot}. Note that this equation for mass variation would be equivalent to $\dot{\ell} = \lambda \ell$ for the longitudinal length if the mass of each section were instead assumed to remain constant.

The system of equations \eqref{eq:lambdadot}, \eqref{eq:omegadot}, \eqref{eq:hddot}, \eqref{eq:deltaddot}, \eqref{eq:vparadot}, and \eqref{eq:rholindot} can be solved in terms of the independent variables $\lambda$, $\Omega$, $H$, $\Delta$, $v_{\parallel}$, and $\Lambda$ as function of time for a given set of initial conditions. Doing this for each section, characterized by a given boundness $\mu$, we can determine the gas properties through the entire stream evolution.

\subsection{Initial conditions}

Consistently with the frozen-in approximation, the initial conditions are set to $\lambda_{\rm i} = \Omega_{\rm i} = \dot{H}_{\rm i} = \dot{\Delta}_{\rm i} = v_{\parallel, \rm i} = 0$. The star is positioned at the tidal radius (see Fig. \ref{fig:sketch}) where $\delta_{\rm i}  = -\pi$ and the initial transverse widths for a given section are $H_{\rm i} = \Delta_{\rm i} = \rstar (1-\mu^2)^{1/2}$. The stellar density profile is assumed to be polytropic with an exponent $\gamma_{\star} = 5/3$. The linear density $\Lambda_{\rm i}$ of each section is obtained by integration over a slice offset by a net distance $r = \mu \rstar$ with respect to the center of the star, following \cite{lodato2009}. The initial pressure is such that hydrostatic equilibrium is satisfied with $\ddot{H}_{\rm p} = -\ddot{H}_{\rm g}$ and $\ddot{\Delta}_{\rm p} = -\ddot{\Delta}_{\rm g}$, which yields a value for the pre-factor $K = 2 \pi^{2/3} G \Lambda_{\rm i}^{1/3} \rstar^{4/3} (1-\mu^2)^{2/3}$.

\section{Results}
\label{sec:evolution}

We now describe the full stream evolution predicted by our model for a fiducial case. We consider a black hole with mass $\mh = 10^6 \msun$ and a star with solar mass and radius $\mstar = \msun$ and $\rstar = \rsun$, which correspond to the typical configuration studied in most past works. 

\subsection{Longitudinal length}

\begin{figure}
\centering
\includegraphics[width=\columnwidth]{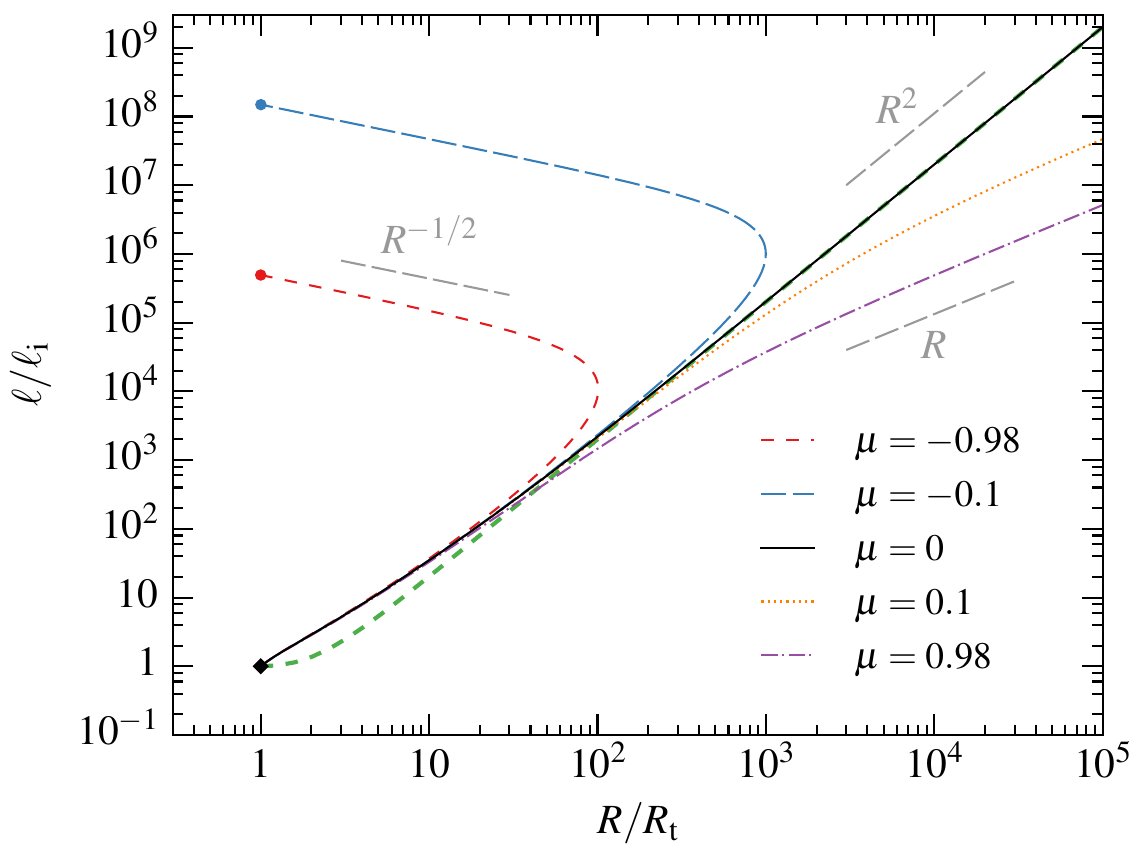}
\includegraphics[width=\columnwidth]{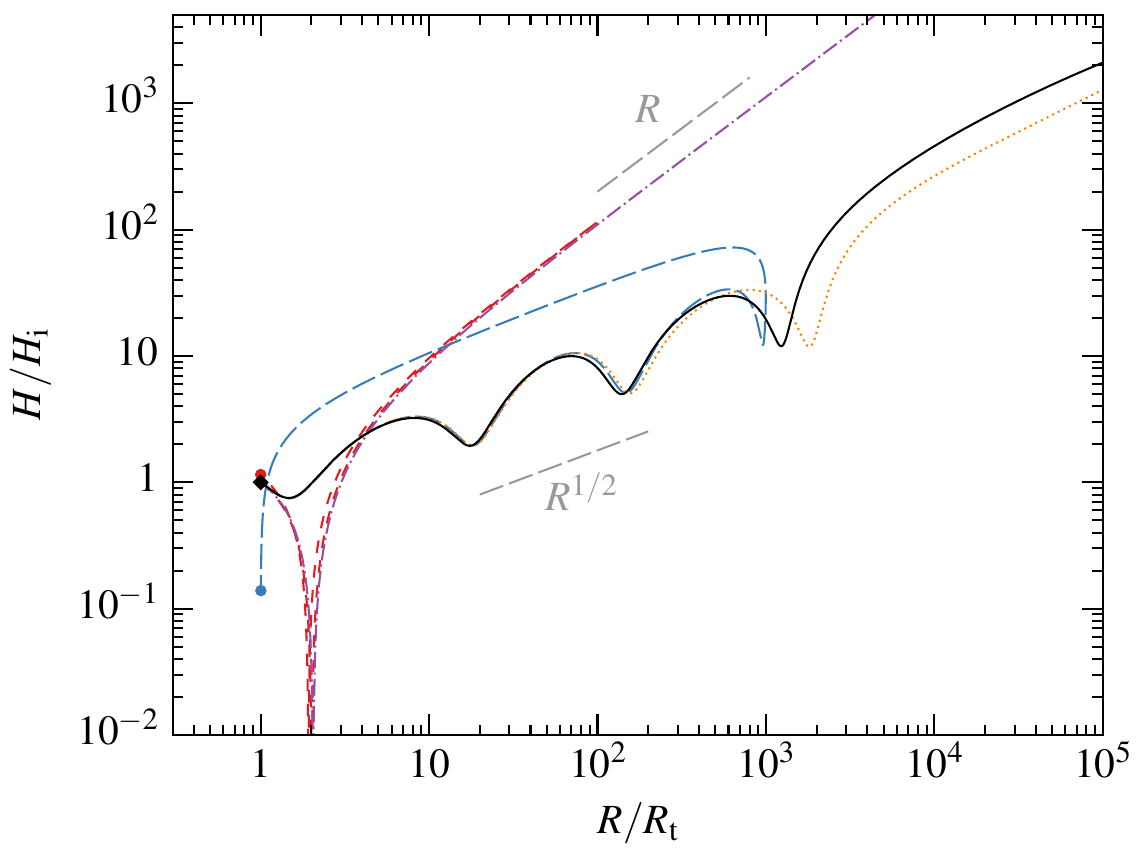}
\includegraphics[width=\columnwidth]{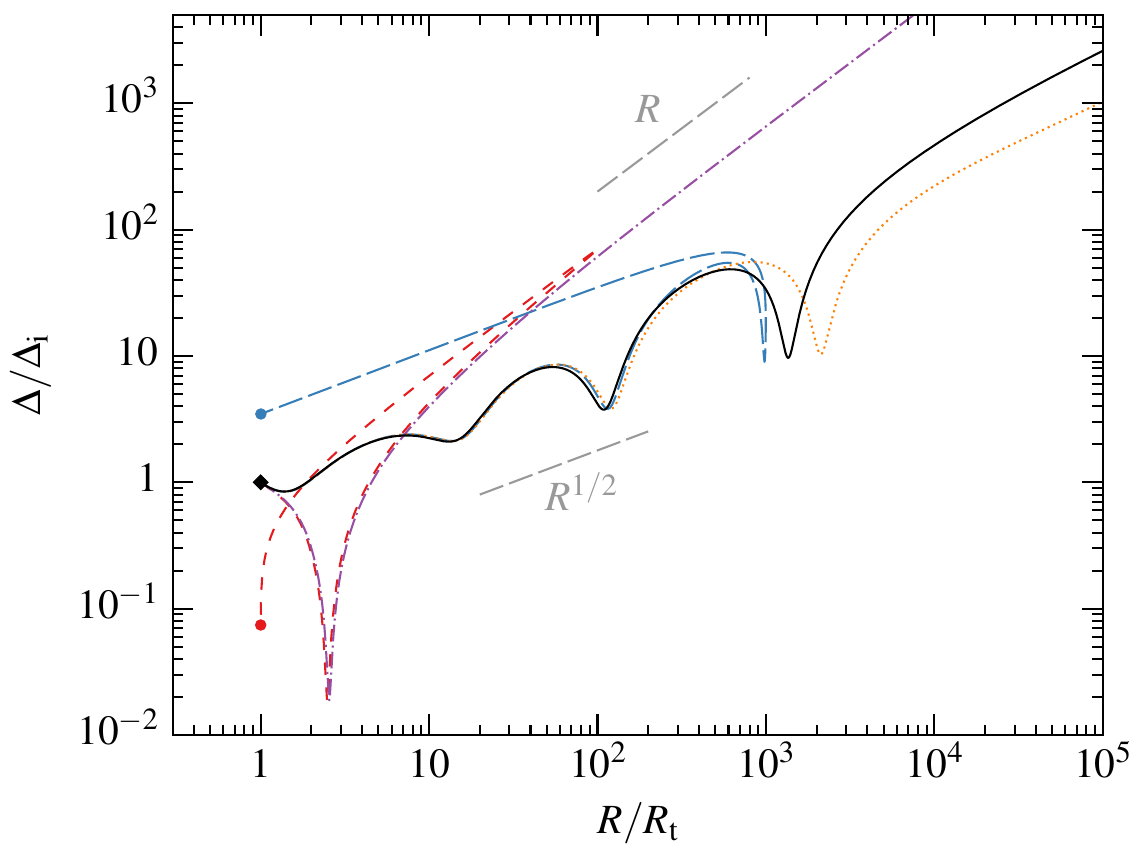}
\caption{Evolution of the longitudinal length (upper panel) and transverse widths in the vertical (middle panel) and in-plane (lower panel) directions for stream sections with different orbital energies as specified by their boundness $\mu$. The initial value used as normalization is arbitrarily set to $\ell_{\rm i}$ for the length, and given by $H_{\rm i} = \Delta_{\rm i} = \rstar (1-\mu^2)^{1/2}$ for the two widths. For a given stream section, diamonds and circles represent initial and final locations, respectively. The thick green dashed line in the upper panel shows the analytical result of equation \eqref{eq:ellana} for $\ell_{\rm eq} = \ell_{\rm i}$ and $R^{\ell}_{\rm eq} = R_{\rm t}$.}
\label{fig:ellhdeltavsr}
\end{figure}

Although the longitudinal length is not solved for explicitly, it can be found from the linear density as $\ell = \ell_{\rm i} (\Lambda/\Lambda_{\rm i})^{-1}$ with an initial value arbitrarily set to $\ell_{\rm i}$. Its evolution is shown as a function of radius in Fig. \ref{fig:ellhdeltavsr} (upper panel) for different stream sections. The sections get stretched with $\dot{\ell} \geq 0$ at all times and irrespective of their boundness $\mu$. When the gas moves outward, the length scales as $\ell \propto R^2$ but this evolution becomes $\ell \propto R$ for the unbound sections with $\mu >0$ because they reach different terminal velocities at large radii. After passing their apocenter, the bound sections with $\mu <0$ move inward with a length increasing as $\ell \propto R^{-1/2}$.

The scaling for the bound sections can be obtained analytically by realizing that, because of their low angular momentum and binding energy, it is legitimate to approximate their center of mass trajectories as radial and parabolic. Equation \eqref{eq:ellddot} then simplifies to\footnote{Note that this equation is equivalent to the system formed by $\dot{\ell} = \lambda \ell$ and equation \eqref{eq:lambdadot} for $\Omega = \delta =0$, which is more convenient to solve for non-radial orbits.} $\ddot{l} = 2 G \mh \ell /R^3$ with $R = (9 G \mh t^2/2)^{1/3}$ \citep{sari2010}. Specifying the initial conditions by $\ell = \ell_{\rm eq}$ and $\dot{\ell} =0$ at $R=R^{\ell}_{\rm eq}$, the length takes the analytical form
\be
\ell = \frac{\ell_{\rm eq}}{5} \left[ 4 \left(\frac{R}{R^{\ell}_{\rm eq}} \right)^{-1/2} + \left( \frac{R}{R^{\ell}_{\rm eq}} \right)^2 \right],
\label{eq:ellana}
\ee
from which we recover the two scalings $\ell \propto R^2$ and $\ell \propto R^{-1/2}$ found above for a stream section when it moves outward and inward, respectively. In particular, the thick green dashed line in Fig. \ref{fig:ellhdeltavsr} shows this analytical formula for $\ell_{\rm eq} = \ell_{\rm i}$ and $R^{\ell}_{\rm eq} = R_{\rm t}$, as imposed by the frozen-in approximation. It is close to the length of the section with $\mu=0$, except near pericenter due to the non-zero gas angular momentum.\footnote{The treatment of stream elongation of \citet{coughlin2016-structure} assuming radial motion is similar to this analytical function. However, they did not include the $\ell \propto R^{-1/2}$ scaling, implying that $\ell = \ell_{\rm i}$ is reached beyond the tidal radius at $R = \sqrt{5} \rt$, which is inconsistent with the frozen-in approximation.}

\begin{figure*}
\centering
\includegraphics[width=0.475\textwidth]{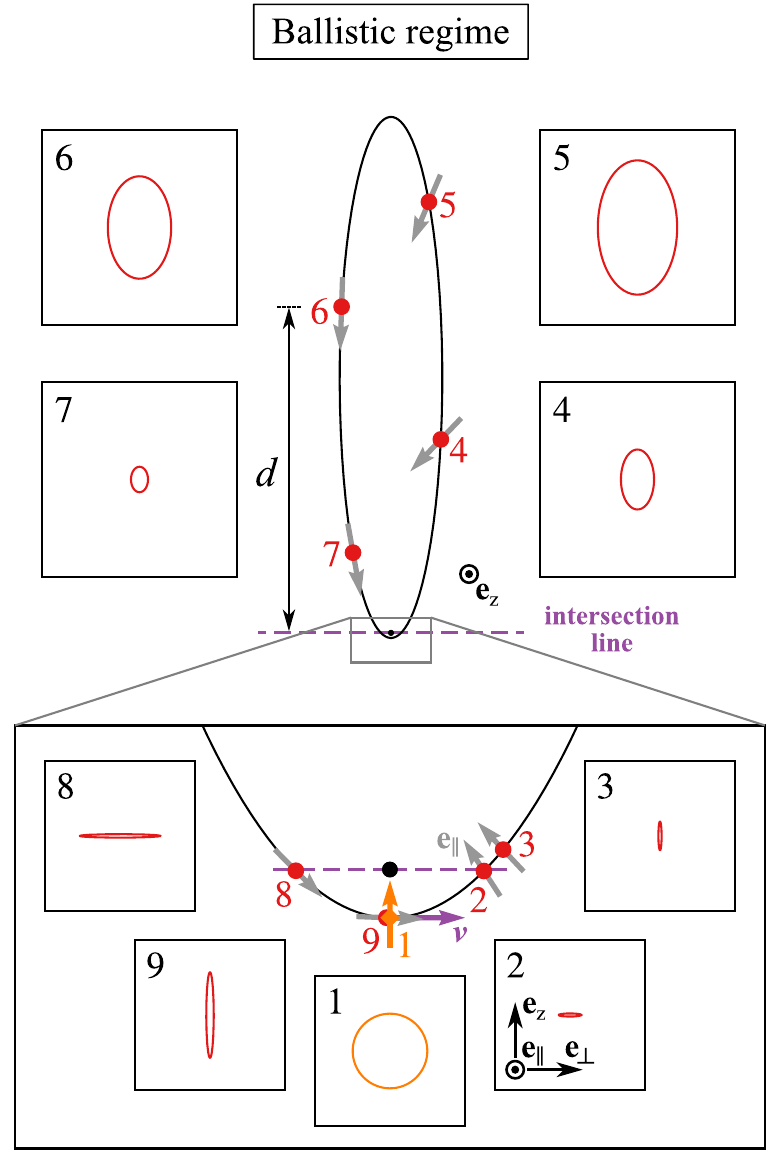}
\hfill
\includegraphics[width=0.475\textwidth]{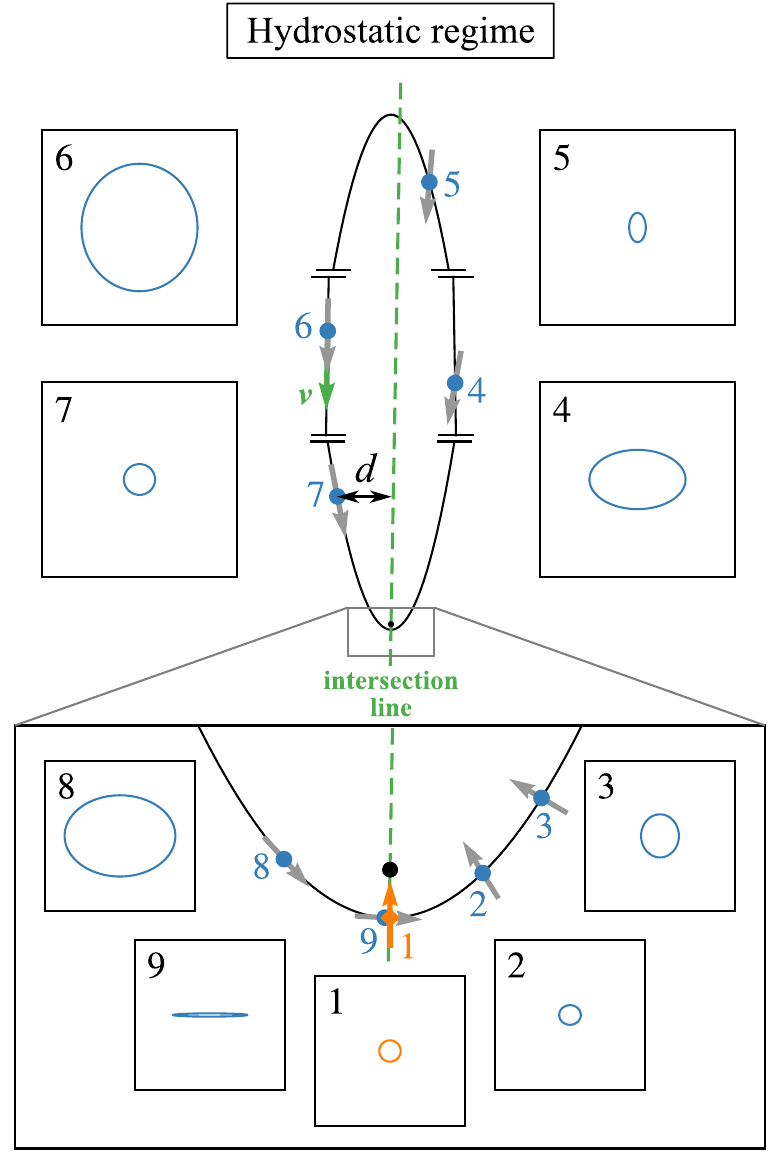}
\vspace{0.2cm}
\caption{Elliptical geometry of two bound stream sections in the ballistic ($\mu = -0.98$, left panel) and hydrostatic ($\mu = -0.1$, right panel) regimes displayed in square boxes at nine different locations (red and blue points) along their center of mass trajectory (black solid curve). The grey arrows represent the unit vector $\vect{e}_{\parallel}$ pointing along the local direction of stream elongation (shown in orange initially). The purple and green arrows show the velocity vector at the equilibrium point where $\dot{H} = 0$, which is parallel to the intersection line (dashed lines of the same color) where the orbital planes of the gas inside each section cross. For better visibility, we only show portions of the trajectory with $\mu = -0.1$ separated by small parallel segments, since it would otherwise have a much larger apocenter. We use the same scale to depict the elliptical sections for boxes inside the lower inset (points 1, 2, 3, 8, and 9), but a different one for the other boxes (points 4, 5, 6, and 7). These scales also differ between the two panels.} 
\label{fig:trajectory}
\end{figure*}

\subsection{Transverse widths}

We now turn to the evolution of the stream sections in the transverse directions. As previously argued by \cite{coughlin2016-structure}, we find that it qualitatively differs depending on the initial location of the section inside the star, which specifies the contribution from the tidal force relative to pressure and self-gravity. The sections belonging to the outer layers have low densities $\rho \ll \rho_{\rm g}  \equiv \mh/(2 \pi R^3)$, implying that their transverse evolution is almost entirely ballistic as specified by the dominant tidal force. Instead, the sections close to the stellar core have higher densities $\rho \gg \rho_{\rm g}$ such that pressure and self-gravity forces dominate to maintain hydrostatic equilibrium at early times. The separation between these two behaviours corresponds to a critical boundness $|\mu| = \mu_{\rm g}$ obtained by setting the initial density to $\rho_{\rm g}$. This leads to $\Lambda_{\rm i} (\mu_{\rm g}) = \mstar (1-\mu^2_{\rm g})/(2 \rstar)$, which numerically gives $\mu_{\rm g} \approx 0.533$ for the polytropic density profile with $\gamma_{\star} = 5/3$ considered here.

We refer to these two regimes as ``ballistic'' ($|\mu|\geq \mu_{\rm g}$) and ``hydrostatic'' ($|\mu|\leq \mu_{\rm g}$). They are discussed separately below based on the evolution of the transverse widths displayed in Fig. \ref{fig:ellhdeltavsr} (middle and lower panels) as a function of radius for different sections. The associated elliptical geometry is shown inside small square boxes in Fig. \ref{fig:trajectory} for two bound sections at nine different locations along the center of mass trajectory.

\subsubsection{Ballistic regime}

The two stream sections with $|\mu| = 0.98$ have low densities $\rho \ll \rho_{\rm g}$ that make their transverse evolution nearly ballistic as specified by the tidal force only \citep{coughlin2016-structure}. The gas with $H>0$ in these sections then follows an almost fixed orbital plane that crosses that of the center of mass along a line. It is straightforward to show geometrically that this intersection line must pass through the black hole and be parallel to the center of mass velocity at the equilibrium point where $\dot{H} = 0$ \citep{luminet1985}.

Due to the frozen-in approximation, this equilibrium point is located at the tidal radius where all sections have an initial velocity vector perpendicular to the direction to the black hole. This implies that the intersection line is aligned with the minor-axis of the orbit (see left panel of Fig. \ref{fig:trajectory}). As a result, the gas in these sections vertically collapses to $H\ll H_{\rm i}$ at the semi-latus rectum where $R \approx 2\rt$ (see also equation 13 of \citealt{stone2013}) before quickly bouncing back due to pressure forces.\footnote{Although we treat them as adiabatic, the collapses experienced by the stream can lead to shocks that increase the gas entropy.} This interaction also occurs during the infall of the bound section, causing the formation of a ``nozzle shock'' \citep{bonnerot2021-nozzle} before (and potentially also after) pericenter passage. Further out, the width is proportional to the projected distance to the intersection line $d \approx R$ since the stream section is confined between the two inclined orbital planes. This implies that $H\propto R$, as seen in Fig. \ref{fig:ellhdeltavsr}.

As we did above for the length, the scaling for the transverse widths of the bound  sections  can  be  obtained analytically by considering a radial parabolic trajectory, which is done by solving $\ddot{H} = \ddot{H}_{\rm t}$ using equation \eqref{eq:hddott} with $R = (9 G \mh t^2/2)^{1/3}$ \citep{sari2010}. With the initial conditions $\dot{H} = 0$ and $H = H_{\rm eq}$ at $R=R^{H}_{\rm eq}$, the (unique) transverse width ($\Delta = H$) is then given by
\be
H = H_{\rm eq} \left[ 2 \left(\frac{R}{R^{H}_{\rm eq}} \right)^{1/2} -  \frac{R}{R^{H}_{\rm eq}}  \right].
\label{eq:hana}
\ee
In the above situation, the equilibrium point is at $R^{H}_{\rm eq} = \rt$, which implies that a collapse with $H=0$ occurs at a radius $R = 4\rt$ only slightly larger than in the non-radial case. The resulting bounce leads to a sign flip in equation \eqref{eq:hana}, such that the same scaling $H \propto R$ as above is followed at larger radii.

A similar collapse occurs along the $\vect{e}_{\perp}$ direction at $R\gtrsim 2\rt$ where $\Delta \ll \Delta_{\rm i}$ (see Figs. \ref{fig:ellhdeltavsr} and \ref{fig:trajectory}), which is the same effect as the in-plane ``pancake'' identified by \cite{coughlin2016-pancakes}. Following a bounce, the width evolves as $\Delta \propto R$ at larger radii like in the vertical direction, which is expected based on equation \eqref{eq:hana} that $\Delta$ also obeys in the radial situation where $\Omega = \delta =0$. For the bound section, the gas gets squeezed again when it comes back near pericenter. This is because the longitudinal direction $\vect{e}_{\parallel}$ (grey arrows in Fig. \ref{fig:trajectory}) gets aligned with the center of mass trajectory. Close to the black hole, $\Delta$ therefore measures the radial extent of the stream section, which is near zero since this gas has similar pericenter distances due to angular momentum conservation under the tidal force alone.

\subsubsection{Hydrostatic regime}

Because the stream sections with $|\mu|\leq 0.1$ have densities $\rho \gg \rho_{\rm g}$ inside the star, the tidal force can initially be neglected compared to self-gravity and pressure. As a result, their transverse widths oscillate around the scaling $H \propto R^{1/2}$ and $\Delta \propto R^{1/2}$ (see Fig. \ref{fig:ellhdeltavsr}) that follows from hydrostatic equilibrium with $\ddot{H}_{\rm p} \approx -\ddot{H}_{\rm g}$ and $\ddot{\Delta}_{\rm p} \approx -\ddot{\Delta}_{\rm g}$ \citep{coughlin2016-structure}. While moving outward, this condition remains satisfied because the density evolves like $\rho_{\rm g}$ as $\rho \propto  R^{-3}$ due to $\Lambda \propto 1/\ell \propto R^{-2}$.

However, the density of the bound section becomes $\rho < \rho_{\rm g} $ after apocenter passage because the scaling changes to $\rho \propto R^{-1/2}$ as $\Lambda \propto R^{1/2}$ during infall. As a result, the tidal force becomes dominant during the approach to the black hole. From this moment, the bound section with $\mu = -0.1$ is confined vertically between the orbital plane of its center of mass and that followed by the gas at $H>0$. Like before, these two planes cross along an intersection line aligned with the velocity vector at the equilibrium point where $\dot{H} = 0$ \citep{luminet1985}. However, this point is now located after apocenter where the gas moves almost radially, implying that the intersection line is nearly aligned with the major-axis of the trajectory (see right panel of Fig. \ref{fig:trajectory}). Because the projected distance then scales as $d \propto R^{1/2}$, so does the vertical width $H$ during the infall, as seen in Fig. \ref{fig:ellhdeltavsr}. Again, this scaling is predicted by equation \eqref{eq:hana} in this case because $R^{H}_{\rm eq} \gg \rt$. This section of stream gets vertically compressed when it passes through the intersection line very close to pericenter, which leads to the ``nozzle shock'' recently simulated by \citet{bonnerot2021-nozzle}.

Because it also obeys equation \eqref{eq:hana} in the radial case, the in-plane width follows the same scaling $\Delta \propto R^{1/2}$ during infall (see Fig. \ref{fig:ellhdeltavsr}). However, its evolution differs near pericenter where it retains a value of $\Delta/\Delta_{\rm i} \approx 4$, larger than for the more bound section evolving under the tidal force alone. The reason is that the additional pressure force induces a spread in angular momentum during the previous phase of hydrostatic equilibrium, which causes the section to acquire pericenter distances different from the stellar center of mass.

\section{Discussion and conclusion}
\label{sec:conclusion}

We have developed a model that can follow the entire evolution of the debris stream by dividing it into individual sections of elliptical geometry. Our model evolves the longitudinal stretching as well as both transverse widths. For the latter, we identified two regimes depending on whether the gas is affected only by the tidal force (“ballistic regime”) or also by pressure and self-gravity (“hydrostatic regime”). By treating the tidal force explicitly and including the gas angular momentum, we are able to accurately follow the dynamics near pericenter and identify the locations where the stream section collapses.

Shortly after its disruption, the star undergoes vertical and in-plane collapses in the ballistic regime that our model can follow simultaneously. While they have so far been investigated independently \citep{stone2013,coughlin2016-pancakes}, our unifying approach can capture the interplay between these two effects. This could be used to more accurately study the hydrodynamics of stellar compression, in particular for deeply-penetrating encounters where most of the gas moves ballistically.

When the stream returns near the black hole, it vertically collapses under the tidal force. In the hydrostatic regime, the gas gets squeezed only once close to pericenter. The resulting nozzle shock was studied in a recent simulation \citep{bonnerot2021-nozzle}, finding that it does not inflate the stream significantly. In the ballistic regime, our model predicts that the stream encounters a vertical collapse both before and after pericenter passage, which we suggest could enhance the impact of the nozzle shock.

After its passage at pericenter, the stream eventually collides with itself, leading to a self-crossing shock that initiates accretion disk formation \citep[e.g.][]{bonnerot2021-light}. Due to a faster vertical width increase and a potentially stronger nozzle shock, our work suggests that earlier-arriving gas in the ballistic regime can collide promptly despite the offset induced by Lense-Thirring precession. Instead, later-arriving gas in the hydrostatic regime is more likely to miss the first collision \citep{bonnerot2021-nozzle} and continue to evolve for several orbital periods before intersecting itself \citep{guillochon2015,batra2021}.\footnote{For our choice of parameters, the stream intersects at a distance $R\approx 100 \rt$ close to the apocenter of the most bound debris. At this location, Fig. \ref{fig:ellhdeltavsr} shows that the vertical width is larger for $\mu = -0.98$ than $\mu = -0.1$, favouring a prompt self-crossing shock in the ballistic regime.} This effect would imply that the accretion flow is fed at a rate that differs from the initial fallback rate, leading to novel observational consequences dependent on black hole spin.

While our present work focuses on a single value of the black hole mass and depth of the encounter, we will carry out a more extensive exploration of the parameter space in the future. A possible effect of a deeper encounter is to modify the fraction of gas that belongs to the hydrostatic and ballistic regimes \citep{steinberg2019}, thus affecting the subsequent stream evolution. Additionally, we intend to generalize our framework to include the presence of stellar rotation and magnetic fields, which may affect the evolution of the transverse widths through the additional centrifugal \citep{golightly2019} and magnetic pressure \citep{bonnerot2017-magnetic,guillochon2017-magnetic} forces acting on the stream.

\begin{acknowledgments}
This project has received funding from the European Union’s Horizon 2020 research and innovation programme under the Marie Sklodowska-Curie grant agreement No 836751. WL is supported by the Lyman Spitzer, Jr. Postdoctoral Fellowship at Princeton University.
\end{acknowledgments}

\bibliography{biblio}{}

\begin{thebibliography}{}
\expandafter\ifx\csname natexlab\endcsname\relax\def\natexlab#1{#1}\fi

\bibitem[{Batra {et~al.}(2021)Batra, Lu, Bonnerot, \& Phinney}]{batra2021}
Batra, G., Lu, W., Bonnerot, C., \& Phinney, E.~S. 2021, ArXiv e-prints, 000,
  arXiv:2112.03918

\bibitem[{Bonnerot \& Lu(2021)}]{bonnerot2021-nozzle}
Bonnerot, C., \& Lu, W. 2021, ArXiv e-prints, arXiv:2106.01376

\bibitem[{Bonnerot {et~al.}(2021)Bonnerot, Lu, \& Hopkins}]{bonnerot2021-light}
Bonnerot, C., Lu, W., \& Hopkins, P.~F. 2021, \mnras, 504, 4885

\bibitem[{Bonnerot {et~al.}(2017)Bonnerot, Price, Lodato, \&
  Rossi}]{bonnerot2017-magnetic}
Bonnerot, C., Price, D.~J., Lodato, G., \& Rossi, E.~M. 2017, \mnras, 469, 4879

\bibitem[{Bricman \& Gomboc(2020)}]{Bricman2020}
Bricman, K., \& Gomboc, A. 2020, \apj, 890, 73

\bibitem[{Coughlin {et~al.}(2016{\natexlab{a}})Coughlin, Nixon, Begelman, \&
  Armitage}]{coughlin2016-structure}
Coughlin, E.~R., Nixon, C., Begelman, M.~C., \& Armitage, P.~J.
  2016{\natexlab{a}}, \mnras, 17, 1

\bibitem[{Coughlin {et~al.}(2016{\natexlab{b}})Coughlin, Nixon, Begelman,
  Armitage, \& Price}]{coughlin2016-pancakes}
Coughlin, E.~R., Nixon, C., Begelman, M.~C., Armitage, P.~J., \& Price, D.~J.
  2016{\natexlab{b}}, \mnras, 455, 3612

\bibitem[{Golightly {et~al.}(2019)Golightly, Coughlin, \&
  Nixon}]{golightly2019}
Golightly, E. C.~A., Coughlin, E.~R., \& Nixon, C.~J. 2019, \apj, 872, 163

\bibitem[{Guillochon \& McCourt(2017)}]{guillochon2017-magnetic}
Guillochon, J., \& McCourt, M. 2017, \apj, 834, L19

\bibitem[{Guillochon \& Ramirez-Ruiz(2015)}]{guillochon2015}
Guillochon, J., \& Ramirez-Ruiz, E. 2015, \apj, 809, 166

\bibitem[{Kochanek(1994)}]{kochanek1994}
Kochanek, C.~S. 1994, \apj, 422, 508

\bibitem[{Lacy {et~al.}(1982)Lacy, Townes, \& Hollenbach}]{lacy1982}
Lacy, J.~H., Townes, C.~H., \& Hollenbach, D.~J. 1982, \apj, 262, 120

\bibitem[{Lodato {et~al.}(2009)Lodato, King, \& Pringle}]{lodato2009}
Lodato, G., King, a.~R., \& Pringle, J.~E. 2009, \mnras, 392, 332

\bibitem[{Luminet \& Marck(1985)}]{luminet1985}
Luminet, J.-P., \& Marck, J.-a. 1985, \mnras, 212, 57

\bibitem[{Rees(1988)}]{rees1988}
Rees, M.~J. 1988, Nature, 333, 523

\bibitem[{Sari {et~al.}(2010)Sari, Kobayashi, \& Rossi}]{sari2010}
Sari, R., Kobayashi, S., \& Rossi, E.~M. 2010, \apj, 708, 605

\bibitem[{Sazonov {et~al.}(2021)Sazonov, Gilfanov, Medvedev, Yao, Khorunzhev,
  Semena, Sunyaev, Burenin, Lyapin, Meshcheryakov, Uskov, Zaznobin, Postnov,
  Dodin, Belinski, Cherepashchuk, Eselevich, Dodonov, Grokhovskaya, Kotov,
  Bikmaev, Zhuchkov, Gumerov, van Velzen, \& Kulkarni}]{sazonov2021}
Sazonov, S., Gilfanov, M., Medvedev, P., {et~al.} 2021, \mnras, 508, 3820

\bibitem[{Steinberg {et~al.}(2019)Steinberg, Coughlin, Stone, \&
  Metzger}]{steinberg2019}
Steinberg, E., Coughlin, E.~R., Stone, N.~C., \& Metzger, B.~D. 2019, Mon. Not.
  R. Astron. Soc. Lett., 485, L146

\bibitem[{Stone {et~al.}(2013)Stone, Sari, \& Loeb}]{stone2013}
Stone, N., Sari, R., \& Loeb, A. 2013, \mnras, 435, 1809

\bibitem[{van Velzen {et~al.}(2021)van Velzen, Gezari, Hammerstein, Roth,
  Frederick, Ward, Hung, Cenko, Stein, Perley, Taggart, Foley, Sollerman,
  Blagorodnova, Andreoni, Bellm, Brinnel, De, Dekany, Feeney, Fremling, Giomi,
  Golkhou, Graham, Ho, Kasliwal, Kilpatrick, Kulkarni, Kupfer, Laher, Mahabal,
  Masci, Miller, Nordin, Riddle, Rusholme, van Santen, Sharma, Shupe, \&
  Soumagnac}]{van_velzen2021}
van Velzen, S., Gezari, S., Hammerstein, E., {et~al.} 2021, \apj, 908, 4

\end{thebibliography}
\bibliographystyle{aasjournal}



\end{document}